\documentclass[lnicst]{svmultln}
\usepackage{times}
\usepackage{graphicx}

\usepackage{epsfig,amsmath,amssymb}
\newcommand{\thm}{\begin{theorem}}
\newcommand{\lem}{\begin{lemma}}
\newcommand{\pro}{\begin{proposition}}
\newcommand{\dfn}{\begin{definition} \rm}
\newcommand{\rem}{\begin{remark}}
\newcommand{\xam}{\begin{example}}
\newcommand{\cor}{\begin{corollary}}
\newcommand{\prf}{\begin{proof}}
\newcommand{\ethm}{\end{theorem}}
\newcommand{\elem}{\end{lemma}}
\newcommand{\epro}{\end{proposition}}
\newcommand{\edfn}{\bbox\end{definition}}
\newcommand{\erem}{\bbox\end{remark}}
\newcommand{\exam}{\bbox\end{example}}
\newcommand{\ecor}{\end{corollary}}
\newcommand{\eprf}{\end{proof}}
\newcommand{\beqn}{\begin{equation}}
\newcommand{\eeqn}{\end{equation}}
\newcommand{\bbox}{\vrule height7pt width4pt depth1pt}
\newcommand{\commentout}[1]{}

\newenvironment{RETHM}[2]{\trivlist \item[\hskip 10pt\hskip\labelsep{\sc #1\hskip 5pt\relax\ref{#2}.}]\it}{\endtrivlist}
\newcommand{\rethm}[1]{\begin{RETHM}{Theorem}{#1}}
\newcommand{\repro}[1]{\begin{RETHM}{Proposition}{#1}}
\newcommand{\relem}[1]{\begin{RETHM}{Lemma}{#1}}
\newcommand{\recor}[1]{\begin{RETHM}{Corollary}{#1}}

\newcommand{\erethm}{\end{RETHM}}
\newcommand{\erepro}{\end{RETHM}}
\newcommand{\erelem}{\end{RETHM}}
\newcommand{\erecor}{\end{RETHM}}

\DeclareMathOperator*{\argmax}{argmax}
\DeclareMathOperator*{\argmin}{argmin}

\begin{document}

\mainmatter

\title{Manipulating Scrip Systems: Sybils and Collusion}

\author{
Ian A. Kash\inst{1} \and
Eric J. Friedman\inst{2} \and
Joseph Y. Halpern\inst{1}
}
\institute{
Computer Science Dept.,
Cornell University,
\email{{kash,halpern}@cs.cornell.edu}
\and
Sch. of Oper. Res.
 and Inf. Eng.,
Cornell University,
\email{ejf27@cornell.edu}
}

\maketitle

\begin{abstract}
Game-theoretic analyses of distributed and peer-to-peer systems
typically use the Nash equilibrium solution concept, but this
explicitly excludes the possibility of strategic behavior involving
more than one agent.  We examine the effects of two types of strategic
behavior involving more than one agent, sybils and collusion, in the
context of scrip systems where agents provide each other with
service in exchange for scrip.  Sybils make an agent more likely to be
chosen to provide service, which generally makes it harder for agents
without sybils to earn money and decreases social welfare.
Surprisingly, in certain circumstances it is possible for sybils to
make all agents better off.
While collusion is generally bad, in the context of scrip systems it
actually tends to
make all agents better off, not merely those who collude.
These results also provide insight into the effects of allowing agents
to advertise and loan money.
While many extensions of
Nash equilibrium have been proposed that address
collusion and other issues relevant to distributed and peer-to-peer
systems, our results show that none of them
adequately address the issues raised by sybils and collusion in scrip
systems.
\end{abstract}

\section{Introduction}\label{sec:intro}

Studies of filesharing networks have shown that more than half of
participants share no files~\cite{adar00,Hughes05}.  Creating a
currency with which users can get paid for the service they provide
gives users an incentive to contribute.  Not surprisingly, scrip
systems have often been proposed to prevent such free-riding,
as well as to address resource-allocation problems more broadly.
For example,
KARMA used scrip to prevent free riding in P2P networks~\cite{karma03}
and Mirage~\cite{mirage} and Egg~\cite{egg} use scrip to allocate
resources in a wireless sensor network testbed and a grid respectively

Chun et al.~\cite{ng05} studied user behavior in a deployed scrip
system and observed that users tried various (rational) manipulations of
the auction mechanism used by the system.  Their observations suggest
that system
designers
will have to deal with game-theoretic concerns.
Game-theoretic analyses of scrip systems (e.g.,
\cite{johari06,scrip06,hens,scrip07}) have focused on Nash
equilibrium.
However, because Nash equilibrium explicitly excludes strategic behavior
involving more than one agent, it cannot deal with many of the
concerns of
systems designers.  One obvious concern is collusion among sets of
agents, but there are more subtle concerns.  In a P2P network, it is
typically easy for an agent to join the system under a number of
different identities, and then mount what has been called
a \emph{sybil attack} \cite{sybil}.  While these concerns are by now
well understood, their impact on a Nash equilibrium is not,
although there has been some work on the effects of multiple
identities in auctions~\cite{yokoo04}.
In this paper we examine the effects of sybils and
collusion on scrip systems.  We show that if such strategic
behavior is not taken into account, the performance of the system can
be significantly degraded; indeed the scrip system can fail
in such a way that
all agents even stop providing
service entirely.  Perhaps more
surprisingly, there are circumstances where sybils and collusion can
improve social welfare. Understanding the circumstances that lead to
these different outcomes is essential to the design of stable and
efficient scrip systems.

In scrip systems where each new user is given an initial amount
of scrip, there is an obvious benefit to creating sybils.
Even if this incentive is removed, sybils are still
useful: they can be used to increase the likelihood that an agent will be
asked to provide service, which makes it easier for him to earn
money.    This means that, in equilibrium, those agents who
have sybils will tend to spend less time without money  and those who
do not will tend to spend more time without money, relative to the
distribution of money if no one had
sybils.  This increases the utility of sybilling agents at the expense
of non-sybilling agents.  The overall effect is such that, if a large
fraction of the agents have sybils (even if each has only a few),
agents without sybils typically will do poorly.
From the perspective of an agent considering creating sybils,
the first few sybils can provide him with a significant benefit, but
the benefits of additional sybils
rapidly diminish.  So if a designer can make sybilling moderately
costly, the number of sybils actually created by rational agents will
usually be relatively small.

If a small fraction of agents have sybils, the situation is more
subtle.  Agents with sybils still do better than those without,
but the situation is not zero-sum.  In particular, changes in the
distribution of money can actually lead to a greater
total number of opportunities to earn money.  This, in turn, can
result in an increase in social welfare: everyone is better off.
However, exploiting this fact is generally not desirable.
The same process that leads to an
improvement in social welfare can also lead to a crash of the system,
where all agents stop providing service.
The system designer can
achieve the same effects  by increasing the average amount of
money or biasing the volunteer selection process, so
exploiting the possibility of sybils is generally not desirable.

Sybils create their effects by increasing the likelihood that an agent will
be asked to provide service;  our analysis of them does not depend
on why this increase occurs.  Thus, our analysis also applies to other
factors that increase demand for an agent's services, such as
advertising.  In particular, our results suggest that there are
tradeoffs involved in allowing advertising.
For example, many systems allow agents to announce their connection
speed and other similar factors.  If this biases requests towards
agents with high connection speeds even when agents with lower
connection speeds are perfectly capable of satisfying a particular
request, then agents with low connection speeds will have a
significantly worse experience in the system.  This also means that
such agents will have a strong incentive to lie about their connection
speed.

While
collusion in generally a bad thing, in the context of
scrip systems with fixed prices, it is almost entirely positive.
Without collusion, if
a user runs out of money he is unable to request service until he is
able to earn some.
However, a colluding group can pool there money so that all members
can make a request whenever the group as a whole has some money.
This increases
welfare for the agents who collude because
agents who have no money receive no service.

Furthermore, collusion tends to benefit the non-colluding agents
as well.  Since colluding agents work less often,
it is easier for everyone to earn money, which ends up making everyone
better off.  However, as with sybils, collusion does
have the potential of crashing the system if the average amount of
money is high.

While a designer should generally encourage collusion, we would expect
that in most systems there will be relatively little collusion and
what collusion exists will involve small numbers of agents.  After
all, scrip systems exist to try and resolve resource-allocation
problems where agents are competing with each other.  If they could
collude to optimally allocate resources within the group, they would
not need a scrip system in the first place.
However, many of the
benefits of collusion come from agents being allowed to effectively
have a negative amount of money (by borrowing from their collusive
partners).  These benefits could also be realized if agents are allowed
to borrow money, so designing a loan mechanism could be an important
improvement for a scrip system.  Of course, implementing such a loan
mechanism in a way that prevents abuse requires a careful design.
These results about sybils and collusion indicate their role in scrip
systems and in distributed and peer-to-peer systems more broadly is
poorly addressed by existing solution concepts.  We conclude by
discussing some of the open questions they raise.

\section{Model}\label{sec:model}

Our model of a scrip system is essentially that of
\cite{scrip07}.  There are $n$ agents in the system.
One agent can request a service which another
agent can volunteer to fulfill.  When a service is performed by agent
$j$ for agent $i$, agent $i$ derives some utility from having that
service performed, while agent $j$ loses some utility for performing
it.  The amount of utility gained by having a service performed and
the
amount lost by performing it may depend on the agent.
We assume that agents have a \emph{type} $t$ drawn from some finite set
$T$ of types.
We can describe the entire system
using the tuple
$(T,\vec{f},n,m)$, where $f_t$ is the fraction with type $t$, $n$ is
the total number of agents, and $m$ is the average amount of money
per agent.
In this paper, we consider only
\emph{standard agents}, whose type we can characterize by a tuple
$t = (\alpha_t, \beta_t, \gamma_t, \delta_t, \rho_t, \chi_t)$, where
\begin{itemize}
\item $\alpha_t$ reflects the cost of satisfying a request;
\item $\beta_t$ is the probability that the agent can satisfy a request
\item $\gamma_t$ measures the utility an agent gains for having a request
satisfied;
\item $\delta_t$ is the rate at which the agents discounts utility (so a
unit of utility in $k$ steps is worth only $\delta^{k/n}_t$ as much as
a unit
of utility now)---intuitively, $\delta_t$ is a measure of an agent's
patience (the larger $\delta_t$ the more patient an agent is, since a
unit of utility tomorrow is worth almost as much as a unit today);
\item $\rho_t$ represents the (relative) request rate (since not all
agents make requests at the same rate)%
---intuitively, $\rho_t$ characterizes an agent's need for service;
and
\item $\chi_t$ represents the relative likelihood of being chosen to
satisfy a request.  This might be because, for example, agents with
better connection speeds are preferred.
\end{itemize}
\noindent The parameter $\chi_t$ did not appear
in~\cite{scrip07}; otherwise the definition of a type is identical to
that of~\cite{scrip07}.

We model the system as running for an infinite number of rounds. In
each round, an agent is picked with probability proportional to
$\rho_t$ to request service: a
particular agent of type $t$ is chosen with
probability $\rho_t / \sum_{t'} f_{t'} n \rho_{t'}$. Receiving
service
costs some amount of scrip that we normalize to \$1. If the chosen
agent does not have enough scrip, nothing will happen in this round.
Otherwise, each agent of type $t$ is able to satisfy this request
with probability $\beta_t$, independent of previous behavior. If at
least one agent is able and willing to satisfy the request, and the
requester has type $t'$, then the requester gets a benefit of
$\gamma_{t'}$ utils (the job is done) and one of the volunteers is
chosen at random (weighted by the $\chi_t$) to fulfill the request.
If the chosen volunteer has type $t$, then that agent loses
$\alpha_t$ utils, and receives a dollar as payment. The utility of
all other agents is unchanged in that round. The total utility of an
agent is the discounted sum of the agent's round utilities. To model
the fact that requests will happen more frequently the more agents
there are, we assume that the time between rounds is $1/n$. This
captures the intuition that things are really happening in parallel
and that adding more agents should not change an agent's request
rate.

In our previous work
\cite{scrip06,scrip07}, we proved a number of results about this model:

\begin{itemize}

\item There is an ($\epsilon$-) Nash equilibrium where each agent
chooses a threshold $k$ and volunteers to work only when he has less
than $k$ dollars.
For this equilibrium, an agent needs no knowledge about other agents,
as long as he knows how often he will make a request and how often he
will be chosen to work (both of which he can determine empirically).

\item Social welfare is essentially proportional to the average number
of agents who have money, which in turn is determined by the types of
agents and the average amount of money the agents in the system have.

\item Social welfare increases as the average amount of money
increases, up to a certain point.  Beyond that point, the system
``crashes'':  the only
equilibrium is the trivial equilibrium where all agents have threshold
0.

\end{itemize}

Our proofs of these results relied on the assumption that all
types of agents shared common values of $\beta$, $\chi$, and $\rho$.
To model sybils and collusion, we need to remove this assumption.
The purpose of the assumption was to make each agent equally likely to
be chosen at each step, which allows entropy to be used to determine
the likelihood of various outcomes, just as in
statistical mechanics~\cite{jaynes}.  When the underlying distribution
is no longer uniform, entropy is no longer sufficient to analyze the
situation, but,
as we show here,
{\em relative entropy}~\cite{cover}
(which can be viewed as a generalization of entropy to allow
non-uniform underlying distributions)
can be used instead.
The essential connection is that where entropy can be interpreted as a
measure of the number of ways a distribution can be realized, relative
entropy can be interpreted as a weighted measure where some outcomes
are more likely to be seen than others.
While this connection is well understood and has been used to derive
similar results for independent random variables~\cite{csiszar84},
we believe
that
our application of the technique to a situation where the
underlying random variables are not independent is novel
and perhaps of independent interest.
Using relative entropy, all our previous results can be extended to
the general case.
The details of this extension are in the appendix.

This model does make a number of simplifying assumptions, but many
can be relaxed without changing the fundamental behavior of the
system (albeit at the cost of greater strategic and analytic
complexity).  For example, rather than all requests having the same
value $\gamma$, the value of a request might be stochastic.  This
would mean that an agent may forgo a low-valued request if he is low
on money.  This fact may impact the threshold he chooses and
introduces new decisions about which requests to make, but the overall
behavior of the system will be essentially unchanged.

A more significant assumption is that prices are fixed.  However,
our results provide insight even if we relax this assumption.  With
variable prices, the behavior of the system depends on the value of
$\beta$.  For large $\beta$, where there are a significant number of
volunteers to satisfy most requests, we expect the resulting
competition to effectively produce a fixed price, so our analysis
applies directly.  For small $\beta$, where there are few volunteers
for each request, variable prices can have a significant impact.
In particular, sybils and collusion are
more likely to result in inflation or deflation rather than a change in
utility.

However, allowing prices to be set endogenously, by bidding, has a number
of negative consequences.  For one thing, it
removes the ability of the system designer to optimize the system using
monetary policy. In addition, for small $\beta$, it makes it possible
for colluding agents to form a cartel to fix prices on a resource they
control.  It also greatly increases the strategic complexity of using
the system: rather than choosing a single threshold, agents need an
entire pricing scheme.
Finally, the search costs and costs of executing a transaction
are likely to be higher with variable prices.
Thus, in many cases we believe that adopting a fixed price or
a small set of fixed prices is a reasonable design decision.

\section{Sybils}\label{sec:sybil}

Unless identities in a system are tied to a real world identity (for
example by a credit card), it is effectively impossible to prevent a
single agent from having multiple identities~\cite{sybil}.
Nevertheless, there are a number of techniques that can make it
relatively costly for an agent to do so.  For example, Credence uses
cryptographic puzzles
to impose a cost each time a new identity wishes to
join the system \cite{credence}.  Given that a designer can impose
moderate costs to sybilling, how much more need she worry about the
problem?  In this section, we show that the gains from creating sybils
when others do not diminish rapidly, so modest costs
may well be sufficient to deter sybilling by typical users.
However, sybilling is a self-reinforcing phenomenon.  As the number
of agents with sybils gets larger, the cost to being a non-sybilling
agent increases and so the incentive to create sybils becomes stronger.
Therefore measures to discourage or prevent sybilling should be taken
early before this reinforcing trend can start.  Finally, we examine the
behavior of systems where only a small fraction of agents have sybils.
We show that under these circumstances a wide variety of outcomes are
possible (even when all agents are of a single type), ranging from a
crash (where no service is provided) to an
increase in social welfare.
This analysis provides insight into the tradeoffs between
efficiency and stability that occur when controlling the money supply
of the system's economy.

\begin{figure}
\begin{minipage}[b]{0.5\linewidth} %
\centering
\includegraphics[height=2.0in]{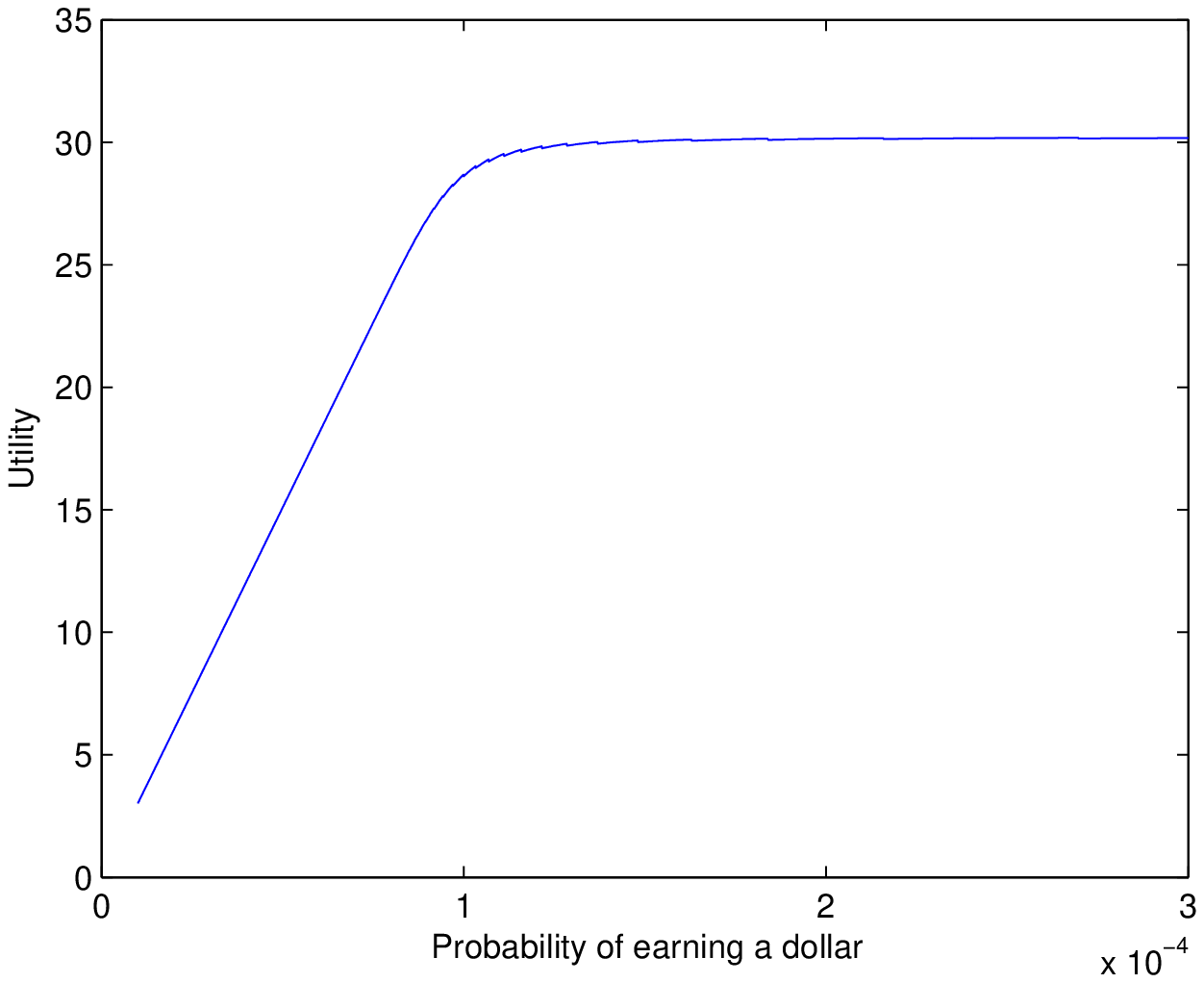}
\caption{The effect of $p_e$ on utility}
\label{fig:sybilben}
\end{minipage}
\hspace{0.5cm} %
\begin{minipage}[b]{0.5\linewidth}
\centering
\includegraphics[height=2.0in]{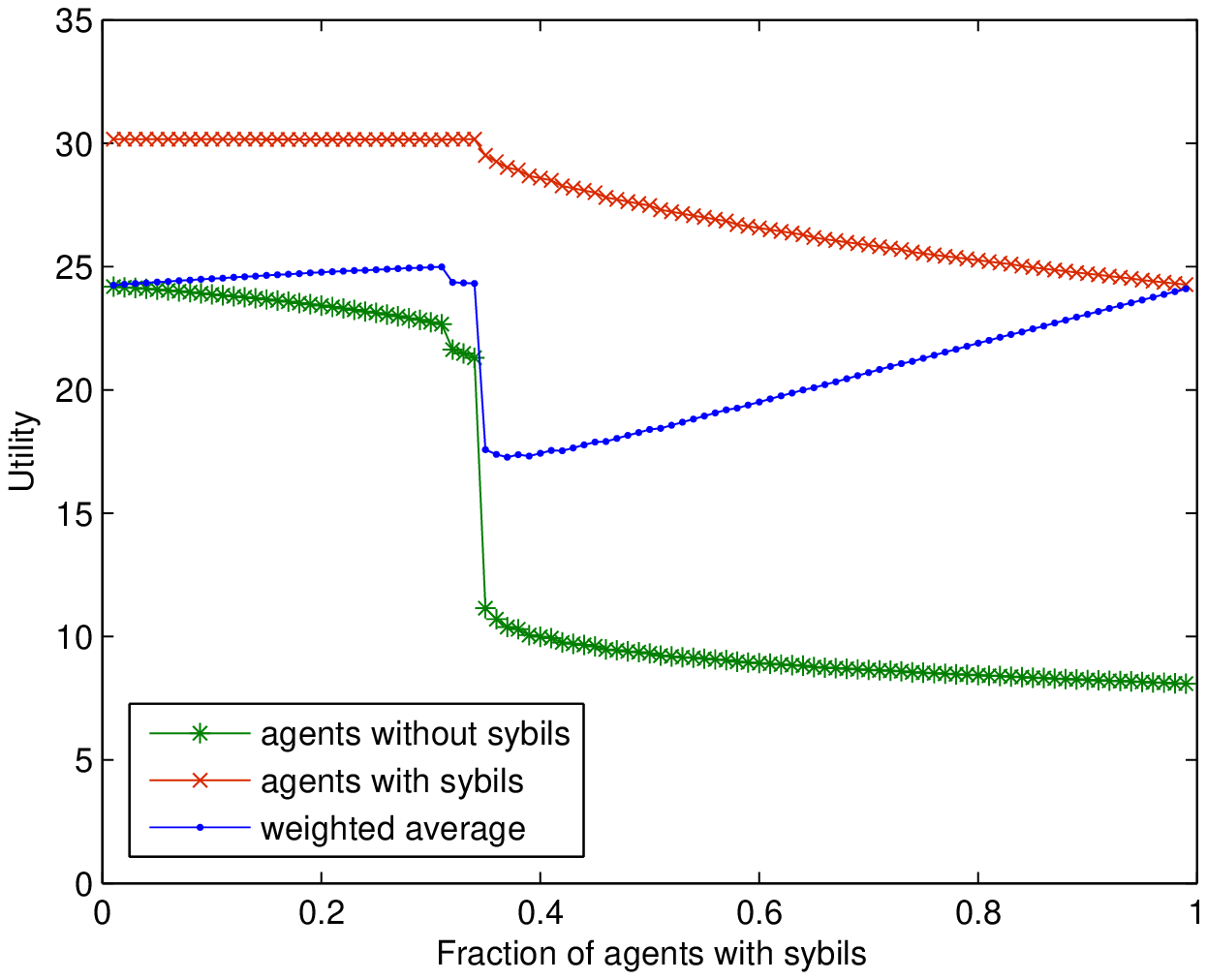}
\caption{The effect of sybils on utility}
\label{fig:sybilcost}
\end{minipage}
\end{figure}

When an agent of type $t$ creates sybils, he does not change anything
about himself but rather how other agents percieve him.  Thus the only
change to his type might be an increase in $\chi_t$ if sybils cause
other agents to choose him more often.  We assume that each sybil is
as likely to be chosen as the original agent, so creating $s$ sybils
increases $\chi_t$ by $s \chi_t$.
(Sybils may have other impacts on the system, such as increased search
costs, but we expect these to be minor.)
The amount of benefit he derives from this depends on two
probabilities: $p_e$ (the probability he will earn a dollar this round
if he volunteers) and $p_s$ (the probability he will be chosen to
spend a dollar and there is another agent willing and able to satisfy
his request).  When $p_e < p_s$, the agent has more opportunities to
spend money than to earn money, so he will regularly have requests go
unsatisfied due to a lack of money.  In this case, the fraction of
requests he has satisfied is roughly $p_e / p_s$, so increasing
$p_e$ results in a roughly linear increase in utility.  When $p_e$ is
close to $p_s$, the increase in satisfied requests is no longer
linear, so the benefit of increasing $p_e$ begins to diminish.
Finally, when $p_e > p_s$, most of the agent's requests are being
satisfied so the benefit from increasing $p_e$ is very small.
Figure~\ref{fig:sybilben} illustrates an agent's utility as $p_e$
varies for $p_s = .0001$.%
\footnote{
Except where otherwise noted, this and other figures assume that $m =
4$, $n = 10000$ and there is
a single type of rational agent with $\alpha = .08$, $\beta = .01$,
$\gamma = 1$, $\delta = .97$, $\rho = 1$, and $\chi = 1$.
These values
are chosen solely for illustration, and are
representative of a broad range of parameter values.
The figures are based on calculations of the equilibrium behavior.
The algorithm used to find the equilibrium is described
in~\cite{scrip07}.}
We formalize the relationship between $p_e$, $p_s$, and utility in the
following theorem.

\thm \label{thm:peps}
Let $p_s$ be the probability that a particular agent is chosen to
make a request in a given round and there is some other agent willing
and able to satisfy it,
$p_e$ be the probability that the agent
earns a dollar given that he volunteers, $r = p_e / p_s$, and  $k$ be
the agent's strategy (i.e., threshold).   In the limit as the number of
rounds goes to
infinity, the fraction of the agent's requests that have an agent
willing and able to satisfy them that get satisfied is
$(r - r^{k+1}) / (1 - r^{k+1})$ if $r \neq 1$ and
$k / (k + 1)$ if $r = 1$.
\ethm

\prf
Since we consider only requests that have another agent willing
and able to satisfy them, the request is satisfied whenever the
agent has a non-zero amount of money.
Since we have a fixed strategy and probabilities, consider the Markov
chain whose states
are  the amount of money the agent has and the transitions describe the
probability of the agent changing from one amount of money to another.
This Markov chain satisfies the requirements to have a stationary
distribution and it can be easily verified that the distribution
gives the agent  probability $r^i (1 - r) / (1 - r^{k+1})$ of having $i$
dollars if $r \neq 1$ and probability $1/(k+1$) if
$r = 1$~\cite{cover}.  This gives
the probabilities given in the theorem.
\eprf

Theorem~\ref{thm:peps} also gives insight into the
equilibrium behavior with sybils.  Clearly, if sybils have no
cost, then creating as many as possible is a dominant strategy.
However, in practice, we expect there is some modest overhead involved
in creating and maintaining a sybil, and that a designer can take steps
to increase this cost without unduly burdening agents.  With such a
cost, adding a sybil might be valuable if $p_e$ is much less than
$p_s$, and a net loss otherwise.  This makes sybils a
self-reinforcing phenomenon.  When a large number of agents create
sybils, agents with no sybils have their $p_e$ significantly
decreased.  This makes
them much worse off and makes sybils much more attractive to them.
Figure~\ref{fig:sybilcost} shows an example of this effect.
This self-reinforcing quality means it is important to take steps to
discourage the use of sybils before they become a problem.  Luckily,
Theorem~\ref{thm:peps} also suggests that a modest cost to create
sybils will often be enough to prevent agents from creating them
because with a well chosen value of $m$, few agents should have low
values of $p_e$.

We have interpreted Figures~\ref{fig:sybilben}~and~\ref{fig:sybilcost}
as being about changes in $\chi$ due to sybils, but the results hold
regardless of what caused differences in $\chi$.  For example, agents
may choose a volunteer based on characteristics such as connection
speed or latency.
If these characteristics are
difficult to verify and do impact decisions, our results show that
agents have a strong incentive to lie about them.
This also suggests that the decision about what sort of information
the system should enable agents to share involves tradeoffs.  If
advertising legitimately allows agents to find better service or more
services they may be interested in, then advertising can increase social
welfare.  But if
these characteristics impact decisions but have little impact on the
actual service, then allowing agents to advertise them can lead to a
situation like that in Figure~\ref{fig:sybilcost}, where some agents
have a significantly worse experience.

\begin{figure}
\begin{minipage}[b]{0.5\linewidth} %
\centering
\includegraphics[height=2.0in]{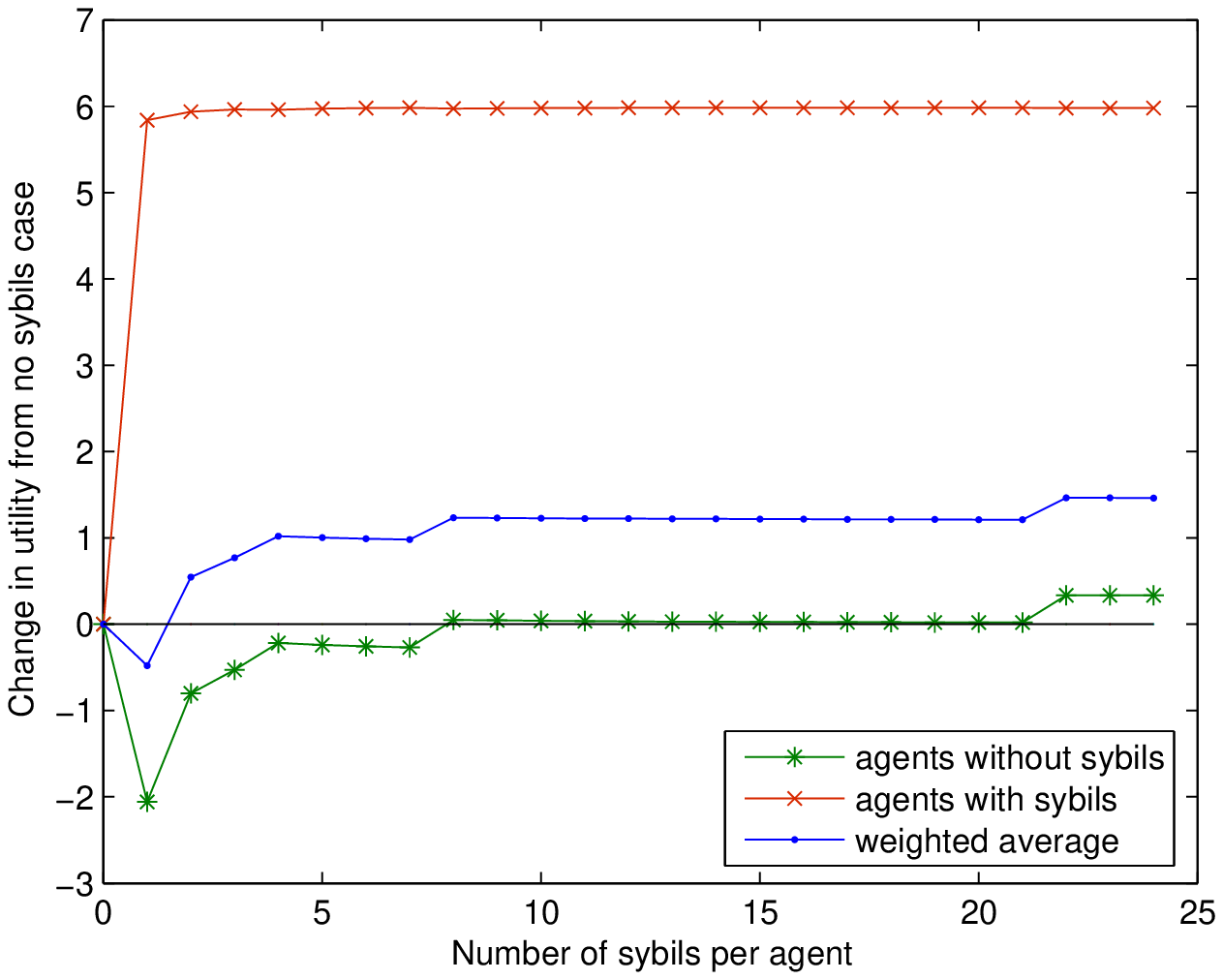}
\caption{Sybils can improve utility}
\label{fig:sybilsmall}
\end{minipage}
\hspace{0.5cm} %
\begin{minipage}[b]{0.5\linewidth}
\centering
\includegraphics[height=2.0in]{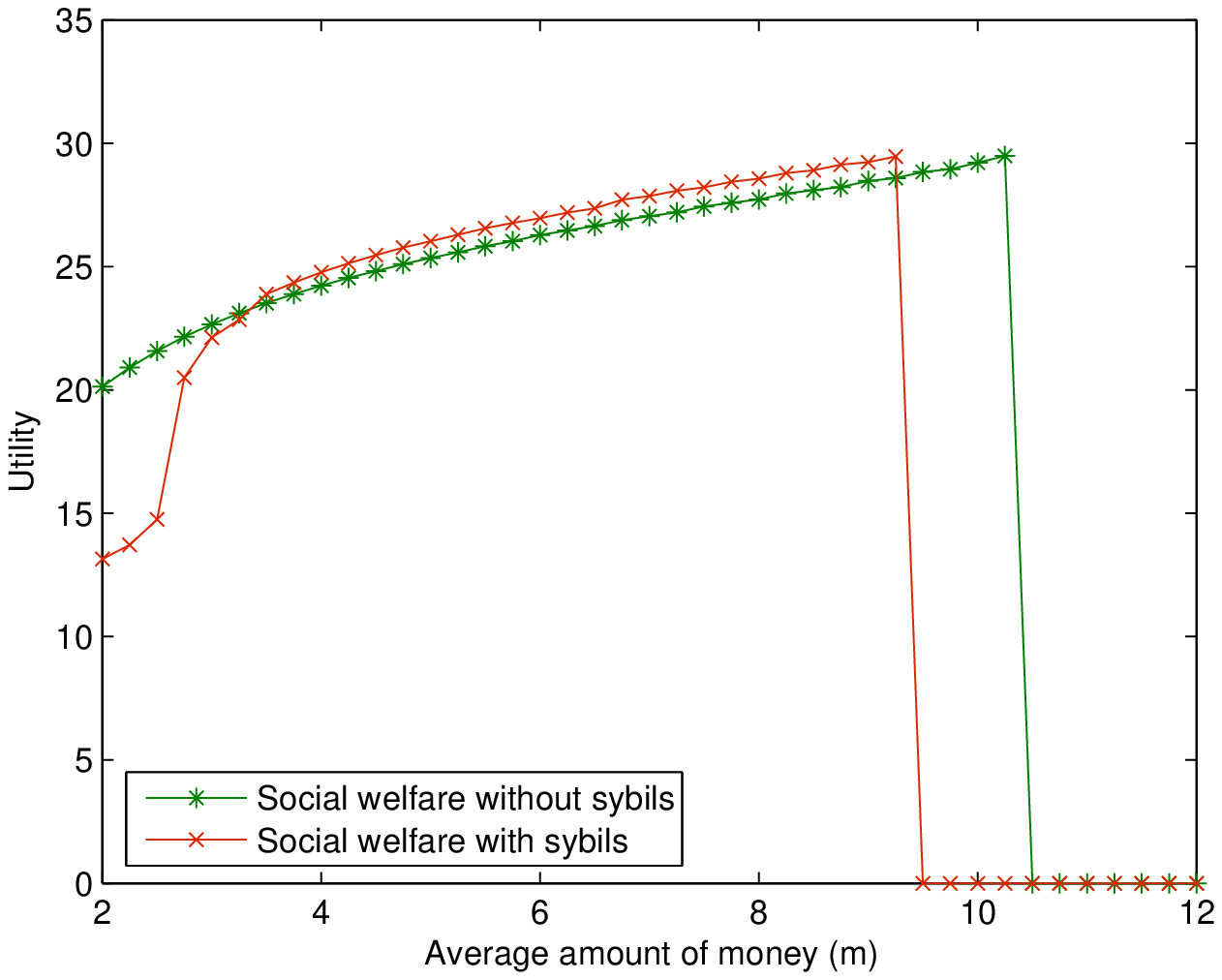}
\caption{Sybils can cause a crash}
\label{fig:sybilcrash}
\end{minipage}
\end{figure}

We have seen that when a large fraction of agents have sybils, those
agents without sybils tend to be starved of opportunities to work.
However, as we saw in Figure~\ref{fig:sybilcost}, when a small
fraction of agents have sybils this effect (and its corresponding
cost) is small.  Surprisingly, if there are few agents with sybils,
an increase in the number of sybils these agents have can actually
result in a \emph{decrease} of their effect on the other agents.
Because agents with sybils are more likely to be chosen to satisfy any
particular request, they are able to use lower thresholds and reach
those thresholds faster than they would without sybils,
so fewer are competing to satisfy any given
request.  Furthermore, since agents with sybils can almost always pay
to make a request, they can provide more opportunities for other agents
to satisfy requests
and earn money.  Social welfare is essentially proportional to
the number of satisfied requests (and is exactly proportional to it if
everyone shares the
same values of $\alpha$ and $\gamma$), so a small number of agents
with a large number of sybils can improve social welfare, as
Figure~\ref{fig:sybilsmall} shows.
Note that this is not necessarily
a Pareto improvement.
For the choice of parameters
in this example, social welfare increases when
the agents create at least two sybils, but agents without sybils are
worse off unless the agents with sybils create at least eight sybils.
As the number of agents with sybils increases, they are forced to
start competing with each other for opportunities to earn money and so
are forced to adopt higher thresholds and this benefit disappears.
This is what causes the discontinuity in Figure~\ref{fig:sybilcost} when
approximately a third of the agents have sybils.

This observation about the discontinuity also suggests another way to
mitigate the negative effects of sybils:
increase the amount of money in the system.
This effect can be seen in Figure~\ref{fig:sybilcrash}, where for $m =
2$ social welfare is very low with sybils but by $m = 4$ it is higher
than it would be without sybils.

Unfortunately, increasing the average amount of money has its own
problems.
Recall from Section~\ref{sec:model} that, if the average amount of money
per agent is too high, the system will crash.
It turns out that just a small number of agents creating
sybils can have the same effect, as
Figure~\ref{fig:sybilcrash} shows.
With no sybils, the point at which social welfare stops increasing and
the system crashes is between $m = 10.25$ and $m = 10.5$.
If one fifth of the agents each create a single sybil, the
system crashes if $m=9.5$,
a point where, without sybils, the social welfare was near optimal.
Thus, if the system designer tries to induce optimal behavior without
taking sybils into account, the system will crash.
Moreover, because of the possibility of a crash,
raising $m$ to tolerate more sybils is  effective only if $m$
was already set conservatively.

This example shows that there is a significant tradeoff between
efficiency and stability.  Setting the money supply high can
increase social welfare, but at the price of making the system less
stable. Moreover, as the following theorem shows, whatever
efficiencies can be achieved with sybils can be achieved without
them, at least if there is only one type of agent. The theorem does
require a technical condition similar in spirit to
{\em $N$-replica economies}~\cite{mascolell} to rule out transient
equilibria that exist only for limited values of $n$.
In our results, we are interested in systems $(T,\vec{f},n,m)$ where
$T$, $\vec{f}$, and $m$ are fixed, but $n$ varies.  This leads to a
small technical problem: there are values of $n$ for which $\vec{f}$
cannot be the fraction of agents of each type nor can $m$ be the
average
amount of money (since, for example, $m$ must be a multiple of
$1/n$).  This technical concern can be remedied in a variety of ways;
the approach we adopt is one used in the literature on $N$-replica
economies

\dfn \label{def:size}
A strategy profile $\vec{k}$ is an \emph{asymptotic equilibrium} for
a system $(T,\vec{f},n,m)$ if for all
$n'$ such that $n'= cn$ for integer $c > 0$,
$\vec{k}$ is a Nash equilibrium for $(T,\vec{f},n',m)$.\edfn

Consider a system where all agents have the same type $t$.
Suppose that some subset of the agents have created sybils, and all the
agents in the subset have created the same number of sybils.  We can
model this by simply taking the agents in the subsets to have a new type
$s$, which is identical to $t$ except that the value of $\chi$ increases.
Thus, we state our results in terms of systems with two types of agents,
$t$ and $s$.

\thm \label{thm:equivalent}
Suppose that $t$ and $s$ are two types that agree except for the value of
$\chi$, and that $\chi_t < \chi_s$.
If $(k_t,k_s)$ is an asymptotic equilibrium for
$(\{t,s\},\vec{f},m)$ with social welfare $x$, then there exists an
$m'$ and $n'$ such that $(k_s)$ is an asymptotic equilibrium for
$(\{ t \}, \{ 1 \}, n', m')$ with social welfare at least $x$. \ethm

\prf
We show this by finding an $m'$ such that agents in the second system
that play some strategy $k$ get essentially the same utility that an
agent with sybils would by playing that strategy in the first system.
Since $k_s$ was the optimal strategy for agents with sybils in the
first system, it must be optimal in the second system as well.
Since agents with sybils have greater utility than those without
(they could always avoid volunteering some fraction of the time to
effectively lower $\chi_s$), social welfare will be
at least as high in the second system as in the first.

Once strategies are fixed, an agent's utility depends only on
$\alpha_t$, $\gamma_t$, $\delta_t$, $p_s$,
 and $p_e$.
  The first three are
constants.  Because the equilibrium is asymptotic, for sufficiently
large $cn$, almost every request by an agent with money is satisfied
(the expected number of agents wishing to volunteer is a constant
fraction of $cn$). Therefore, $p_s$ is essentially $1/cn$, his
probability of being chosen to make a request.  With fixed
strategies, any value of $p_e$ between 0 and $\beta_t$ can be
achieved by taking the appropriate $m'$.  Take $m'$ such that, if
every agent in the second system plays $k_s$, the resulting $p_e$
will be the same as the agents with sybils had in the original
equilibrium. Note that $p_s$ may decrease slightly because fewer
agents will be willing to volunteer, but we can take $n' = cn$ for
sufficiently large $c$ to make this decrease arbitrarily small. \eprf

The analogous result for systems with more than one type of agent is
not true.  Consider the situation shown in
Figure~\ref{fig:sybilcost}, where forty percent of the agents have
two sybils.  With this population, social welfare is lower than if
no agents had sybils. However, the same population could be
interpreted simply as having two different types, one of whom is
naturally more likely to be chosen to satisfy a request.  In this
situation, if the agents less likely to be chosen created exactly
two sybils each, the each agent would then be equally likely to be
chosen and social welfare would increase.
While changing $m$ can change the relative quality of the two
situations, a careful analysis of the proof of
Theorem~\ref{thm:equivalent} shows that, when each population is
compared using its optimal value of $m$, social welfare is greater
with sybils.
While situations like this
show that it is theoretically possible for sybils to increase social
welfare beyond what is possible by adjusting the average amount of
money, this outcome seems unlikely in practice. It relies on
agents creating just the right number of sybils. For situations
where such a precise use of sybils would lead to a significant
increase in social welfare, a designer could instead improve social
welfare by biasing the algorithm agents use for selecting which
volunteer will satisfy the request.

\section{Collusion}\label{sec:collusion}

Agents that collude gain two major benefits.  The primary benefit is
that they can share money, which simultaneously makes them less likely
to run out of money and be unable to make a request and allows them to
pursue a joint strategy for determining when to work.  The secondary
benefit, but important in particular for larger collusive groups, is
that they can satisfy each other's requests.  The effects of collusion
on the rest of the system depend crucially on whether agents are
able to volunteer to satisfy requests when they personally cannot
satisfy the request but one of their colluding partners can.  In a
system where a request is for computation, it seems relatively
straightforward for an agent to pass the computation to a partner to
perform and then pass the answer back to the requester.  On the other
hand, if a request is a piece of a file it seems less plausible that
an agent would accept a download from someone other than the person he
expects and it seems wasteful to have the chosen volunteer download it
for the sole purpose of immediately uploading it.  If it is possible
for colluders to pass off requests in this fashion, they are able to
effectively act as sybils for each other, with all the consequences we
discussed in Section~\ref{sec:sybil}.  However, if agents can
volunteer only for requests they can personally satisfy, the effects of
collusion are almost entirely positive.

Since we have already discussed the consequences of sybils, we will
assume that agents are able to volunteer only to satisfy requests that
they personally can satisfy.
Furthermore, we make the simplifying assumption that agents that
collude are of the same type, because if agents of different types
collude their strategic decisions become more complicated.  For
example, once the colluding group has accumulated a certain amount of
money it may wish to have only members with small values of
$\alpha$ volunteer to satisfy requests.  Or when it is low on money it
may wish to deny use of money to members with low values of $\gamma$.
This results in strategies that involve sets of thresholds rather than
a single threshold, and while nothing seems fundamentally different
about the situation, it makes calculations significantly more difficult.

With these assumptions, we now examine how colluding agents will
behave.
Because colluding agents share money and types, it is irrelevant which
members actually perform work and have money.  All that matters is the
total amount of money the group has.  This means that when the group
needs money, everyone in the group volunteers for a job.  Otherwise no
one does.  Thus, the group essentially acts like a single
agent, using a threshold
which will be
somewhat less than the sum of the thresholds
that the individual agents
would have used,
because it is less likely that $c$ agents will make $ck$ requests in
rapid succession than a single agent making $k$.  Furthermore, some
requests will not require scrip at all because they can potentially be
satisfied by other members of the colluding group.
When deciding whether the
group should satisfy a member's request or ask for an outside
volunteer to fulfill it, the group must decide whether it should pay a
cost of $\alpha$ to avoid spending a dollar.  Since not spending a
dollar is effectively the same as earning a dollar, the decision is
already optimized by the threshold strategy; the group should always
attempt to satisfy a request internally unless it is in a temporary
situation where the group is above its threshold.

\begin{figure}[htb]
\centering \epsfig{file=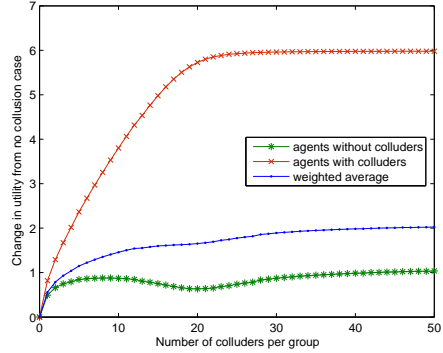, height=2.0in}
\caption{The effect of collusion on utility}
\label{fig:numcollusion}
\end{figure}

Figure~\ref{fig:numcollusion} shows an example of the effects of
collusion on agents' utilities as the size of collusive groups
increases.  As this figure suggests, the effects typically go through
three phases.  Initially, the fraction of requests colluders satisfy
for each other is small.  This means that each collusive group must
work for others to pay for almost every request its members make.
However, since they share money, the colluders do not have to work as
often as individuals would.  Thus, other agents have more
opportunity to work, and every agent's $p_e$ increases,
making all agents better off.

As the number of colluders increases,
the fraction of requests they satisfy internally grows significant.
We can think of $p_s$ as decreasing in this case, and view these
requests as being satisfied ``outside'' the scrip system because no
scrip changes hands.
This is good for colluders,
but is bad for other agents whose $p_e$ is lower
since fewer requests are being made.  Even in this range,
non-colluding agents still tend to be better off than if
there were no colluders because the overall competition for
opportunities to work is still lower.  Finally, once the collusive
group is large enough, it will have a low $p_s$ relative to $p_e$.
This means the collusive group can use a very low threshold which
again begins improving utility for all agents.  The equivalent
situation with sybils was transitory and disappeared when more agents
created sybils.
However,
this low threshold is an inherent
consequence of colluders satisfying each other's requests, and so
persists and even increases as the amount of collusion in the system
increases.
Since collusion is difficult to maintain (the problem of incentivizing
agents to contribute is the whole point of using scrip), we would
expect the size of collusive groups seen in practice to be relatively
small.  Therefore, we expect that for most systems collusion will be
Pareto improving.  Note that, as with sybils, this decreased
competition can also lead to a crash.  However, if the system designer
is monitoring the system and encouraging and expecting collusion she
can reduce $m$ appropriately and prevent a crash.

These results also suggest that creating the
ability to take out loans (with an appropriate interest rate) is
likely to be beneficial.
Loans gain the benefits of reduced competition without the
accompanying cost of fewer requests being made in the system.
However, implementing a loan mechanism requires addressing a number of
other incentive problems.  For example, {\em whitewashing}, where
agents take on a new identity (in this case to escape debts) needs to
be prevented~\cite{FrR01}.

\section{Conclusion}\label{sec:conclusion}

We have seen that sybils are generally bad for social welfare in scrip
systems, and that where they are beneficial the same results can be
achieved without sybils by increasing the amount of money or biasing
volunteer selection in the
system.  On the other hand, collusion tends to be a net benefit and
should be encouraged.  Indeed, the entire purpose of the system is to
allow users to collude and provide each other with service despite
incentives to free ride.  Beyond the context of scrip systems, these
results raise questions about game-theoretic analysis of distributed
and peer-to-peer systems more generally.

Many solution concepts that consider joint strategic behavior have
been proposed.
Solution concepts that require that no collusive group have an
incentive to deviate,
such as \emph{strong} Nash equilibrium~\cite{strong} and
\emph{coalition-proof} Nash equilibrium~\cite{cpne},
are well known.
More recently, the possibility that colluding agents might be
malicious or simply behave in an unpredictable fashion has been
considered.  $k$-fault tolerant Nash equilibrium requires that a each
agent's strategy is still optimal even if up to $k$ agents
behave in an arbitrary fashion~\cite{eliaz02}. The BAR model has three
types of agents: Byzantine, rational, and obedient ``altruistic''
agents who will follow the given protocol even if it is not
rational~\cite{bargames,bargossip}.
$(k,t)$-robustness
combines both these directions with the possibility that $k$ agents
may behave arbitrarily and $t$ agents may collude~\cite{ADGH06}.

The possibility of
creating sybils has received
significantly less attention and is more unique to distributed and
peer-to-peer systems.
Sybils
can have a tremendous impact on the efficiency of the system and
understanding the incentives involved in creating them is crucial to
managing them appropriately.  Finding a solution concept that accounts
for sybils in a rigorous fashion is still an open question.

Collusion is beneficial for most scrip systems, but in a way that is
poorly suited for most solution concepts that address it. Existing
solution concepts look for strategies that remain stable in the
presence of collusion.  For example, $k-t$ robustness allows $t$
agents to collude and then requires that other agents have no
incentive to change their strategy.  Our results have shown
that agents do have an incentive to change their strategies in
response, which means that no equilibrium exists according to these
solution concepts.  While the particular equilibrium strategies
change, scrip systems are still stable (and in fact improved) in the
presence of collusion.  Existing solution concepts that address
collusion are inadequate in cases where collusion is beneficial.

\bibliographystyle{abbrv}
\bibliography{Z:/Research/Bibliography/kash}

\begin{thebibliography}{10}

\bibitem{ADGH06}
I.~Abraham, D.~Dolev, R.~Gonen, and J.~Halpern.
\newblock Distributed computing meets game theory: Robust mechanisms for
  rational secret sharing and multiparty computation.
\newblock In {\em Proc.~25th ACMSymp.~on Principles of Distributed Computing
  ({PODC})}, pages 53--62, 2006.

\bibitem{adar00}
E.~Adar and B.~A. Huberman.
\newblock Free riding on {G}nutella.
\newblock {\em First Monday}, 5(10), 2000.

\bibitem{johari06}
C.~Aperjis and R.~Johari.
\newblock A peer-to-peer system as an exchange economy.
\newblock In {\em GameNets '06: Proceeding from the 2006 Workshop on Game
  Theory for Communications and Networks}, page~10, 2006.

\bibitem{strong}
R.~J. Aumann.
\newblock Acceptable points in general cooperative $n$-person games.
\newblock In A.~Tucker and R.~Luce, editors, {\em Contributions to the Theory
  of Games IV, Annals of Mathematical Studies 40}, pages 287--324. Princeton
  University Press, Princeton, N. J., 1959.

\bibitem{cpne}
D.~Bernheim, B.~Peleg, and M.~Whinston.
\newblock Coalition-proof {N}ash equilibria {I}. concepts.
\newblock {\em Journal of Economic Theory}, 42:1--12, 1987.

\bibitem{egg}
J.~Brunelle, P.~Hurst, J.~Huth, L.~Kang, C.~Ng, D.~Parkes, M.~Seltzer,
  J.~Shank, and S.~Youssef.
\newblock Egg: An extensible and economics-inspired open grid computing
  platform.
\newblock In {\em Third Workshop on Grid Economics and Business Models
  ({GECON})}, pages 140--150, 2006.

\bibitem{mirage}
B.~Chun, P.~Buonadonna, A.~AuYung, C.~Ng, D.~Parkes, J.~Schneidman, A.~Snoeren,
  and A.~Vahdat.
\newblock Mirage: A microeconomic resource allocation system for sensornet
  testbeds.
\newblock In {\em Second IEEE Workshop on Embedded Networked Sensors}, pages
  19--28, 2005.

\bibitem{bargames}
A.~Clement, J.~Napper, H.~C. Li, J.-P. Martin, L.~Alvisi, and M.~Dahlin.
\newblock Theory of {BAR} games.
\newblock In {\em Proc.~26th ACM Symp.~on Principles of Distributed Computing
  ({PODC})}, pages 358--359, 2007.

\bibitem{cover}
T.~Cover and J.~Thomas.
\newblock {\em Elements of Information Theory}.
\newblock John Wiley {\&} Sons, Inc., New York, 1991.

\bibitem{csiszar84}
I.~Csisz{\'a}r.
\newblock Sanov propery, generalized i-projection and a conditional limit
  theorem.
\newblock {\em The Annals of Probability}, 12(3):768--793, 1984.

\bibitem{sybil}
J.~R. Douceur.
\newblock The sybil attack.
\newblock In {\em First International Workshop on Peer-to-Peer Systems
  (IPTPS)}, pages 251--260, 2002.

\bibitem{eliaz02}
K.~Eliaz.
\newblock Fault tolerant implementation.
\newblock {\em Review of Economic Studies}, 69:589--610, 2002.

\bibitem{scrip06}
E.~J. Friedman, J.~Y. Halpern, and I.~A. Kash.
\newblock Efficiency and {N}ash equilibria in a scrip system for {P2P}
  networks.
\newblock In {\em Proc.~Seventh ACM Conference on Electronic Commerce (EC)},
  pages 140--149, 2006.

\bibitem{FrR01}
E.~J. Friedman and P.~Resnick.
\newblock The social cost of cheap pseudonyms.
\newblock {\em Journal of Economics and Management Strategy}, 10(2):173--199,
  2001.

\bibitem{MaxEnt}
A.~J. Grove, J.~Y. Halpern, and D.~Koller.
\newblock Random worlds and maximum entropy.
\newblock {\em J. Artif. Intell. Res. (JAIR)}, 2:33--88, 1994.

\bibitem{hens}
T.~Hens, K.~R. Schenk-Hoppe, and B.~Vogt.
\newblock The great {C}apitol {H}ill baby sitting co-op: Anecdote or evidence
  for the optimum quantity of money?
\newblock {\em J. of Money, Credit and Banking}, 9(6):1305--1333, 2007.

\bibitem{Hughes05}
D.~Hughes, G.~Coulson, and J.~Walkerdine.
\newblock Free riding on {G}nutella revisited: The bell tolls?
\newblock {\em IEEE Distributed Systems Online}, 6(6), 2005.

\bibitem{jaynes}
E.~T. Jaynes.
\newblock Where do we stand on maximum entropy?
\newblock In R.~D. Levine and M.~Tribus, editors, {\em The Maximum Entropy
  Formalism}, pages 15--118. MIT Press, Cambridge, Mass., 1978.

\bibitem{scrip07}
I.~A. Kash, E.~J. Friedman, and J.~Y. Halpern.
\newblock Optimizing scrip systems: Efficiency, crashes, hoarders and
  altruists.
\newblock In {\em Proc.~Eighth ACM Conference on Electronic Commerce (EC)},
  pages 305--315, 2007.

\bibitem{bargossip}
H.~C. Li, A.~Clement, E.~L. Wong, J.~Napper, I.~Roy, L.~Alvisi, and M.~Dahlin.
\newblock {BAR} gossip.
\newblock In {\em Sixth Symp.~on Operating Systems Design and Implementation
  ({OSDI})}, pages 191--204, 2006.

\bibitem{mascolell}
A.~Mas-Colell, M.~D. Whinston, and J.~R. Green.
\newblock {\em Microeconomic Theory}.
\newblock Oxford University Press, Oxford, U.K., 1995.

\bibitem{ng05}
C.~Ng, P.~Buonadonna, B.~Chun, A.~Snoeren, and A.~Vahdat.
\newblock Addressing strategic behavior in a deployed microeconomic resource
  allocator.
\newblock In {\em Third Workshop on Economics of Peer-to-Peer Systems
  ({P2PECON})}, pages 99--104, 2005.

\bibitem{karma03}
V.~Vishnumurthy, S.~Chandrakumar, and E.~G. Sirer.
\newblock {KARMA:} a secure economic framework for peer-to-peer resource
  sharing.
\newblock In {\em First Workshop on Economics of Peer-to-Peer Systems
  (P2PECON)}, 2003.

\bibitem{credence}
K.~Walsh and E.~G. Sirer.
\newblock Experience with an object reputation system for peer-to-peer
  filesharing.
\newblock In {\em Third Symp.~on Network Systems Design \& Implementation
  ({NSDI})}, pages 1--14, 2006.

\bibitem{yokoo04}
M.~Yokoo, Y.~Sakurai, and S.~Matsubara.
\newblock The effect of false-name bids in combinatorial auctions: new fraud in
  internet auctions.
\newblock {\em Games and Economic Behavior}, 46(1):174--188, 2004.

\end{thebibliography}

\appendix

\section{Using Relative Entropy}\label{sec:relent}

Our previous results showed the existence of equilibria in a simple
class of strategy called threshold strategies. The threshold
strategy $S_k$ is the strategy where an agent volunteers if and only
if his current amount of money is less than $k$.  This class of
strategies is attractive for several reasons.  From the perspective
of an agent, it makes his decisions simple; he requires no
complicated computation or state to determine whether to volunteer.
It is also quite natural.  If an agent is feeling poor (less than
$k$) he will want to work to earn money.  If an agent feels rich (at
least $k$) he would rather defer the costs associated with working
into the future.  From the perspective of the design and analysis of
scrip systems, these strategies are nice because they require little
information to determine what will happen.  An agent's decision is
determined solely by his threshold and current amount of money.  This
allows for straightforward evaluation of the evolution of the system
as a Markov chain and of agent's strategic decisions as a Markov
decision process (MDP).

Our theoretical results follow from three key facts about the
Markov Chain and MDP:
\begin{enumerate}

\item The Markov chain that describes the system has a ``well
behaved'' limit distribution when all agents play threshold
strategies (Lemma~3.1~of~\cite{scrip07}).

\item There is a distribution of money $d^*$ (the fraction of agents
with each amount of money) such that, after some initial period, the
observed distribution of money in the system will almost always be
``close'' to $d^*$ (Theorem~3.1~of~\cite{scrip07}).

\item If some agents raise their thresholds and all other agents keep
their thresholds unchanged then the new distribution $d^+$ will have
at least as large a fraction with zero dollars and at most as large a
fraction at their threshold amount of money as $d^*$
(Lemma~3.2~of~\cite{scrip07}).

\end{enumerate}

Once we abandon the assumption of payoff heterogeneity, these facts
remain true, but we have to modify our arguments for them.  First, the
limit distribution is still well
behaved.  Suppose agent $i$ is twice as likely to spend money as
agent $j$ and consider two states which are identical except that in
one agent $i$ has one dollar and agent $j$ has none and while in the
other the situation is reversed.  Since agent $i$ spends money
faster once he gets it, we would expect it would be less likely to
see him holding a dollar than agent $j$.  In fact, the second state
turns out to be exactly twice as likely as the first.  Using this
observation we can build a simple weighting of the relative
likelihood of states based on the division of money among agents
(see Lemma~\ref{lem:stationary}).

The next fact is that the distribution of money in the system will
almost always be close to a particular distribution $d^*$.  This
type of clustering about the most likely distribution is common in
statistical physics (for example the arrangement of molecules of a
gas in a closed container), where it is known as a {\em
concentration phenomenon}~\cite{jaynes}.  Because each possible
arrangement is equally likely, the observed outcomes are
concentrated around the distribution that maximizes entropy.  Our
states are not equally likely, but we do have a weighting that
captures the relative likelihoods.  We can use this weighing through
a generalization of entropy known as {\em relative
entropy}~\cite{cover}.  The relative entropy can be thought of as a
distance measure for distributions.  For distributions $p$ and $q$,
the relative entropy is $D(p~||~q) = \sum_{x} p(x) \log (p(x) /
q(x))$.  Finding the distribution that maximizes entropy is a
special case of relative entropy: it is the distribution that
minimizes relative entropy relative to the uniform distribution. The
$d^*$ we want is the distribution the minimizes relative entropy
relative to a distribution based on the limit distribution of the
Markov chain. The formula for this distribution is given by
Lemma~\ref{lem:minrelent} and the argument that there is a
concentration phenomenon for it is the subject of
Lemma~\ref{lem:relent}.  The importance of relative entropy to
concentration phenomena is examined in much greater generality by
Csisz\'ar~\cite{csiszar84}.
However, those results cover independent random variables that are
made dependent by constraints (for example a series of tosses of a
biased coin with the constraint that at least 75 percent were
heads). We believe our application of the technique to the amounts
of money agents have, which are not independent, is novel
and perhaps of independent interest.

The final fact involves using this formula to show that this
distribution has natural monotonicity properties: if some agents
increase their threshold, then
there will be more agents with no money and fewer agents at their
threshold.  Intuitively, this is because more of those agents will want
to work at any given time so money is harder to earn.  When money is
harder to earn but agents want to spend as often, agents will find it
harder to reach their threshold amount of money and spend more time
with no money.  This intuition is formalized in
Lemma~\ref{lem:monotone}.

These lemmas are sufficient to generalize our previous results.  In
particular, this means

\begin{itemize}

\item when all agents use threshold strategies, each has an
($\epsilon$-)~best response that is a threshold strategy;

\item these best responses are non-decreasing in the strategies of the
other agents; and as a consequence

\item ($\epsilon$-)~Nash equilibria exist where all agents play
threshold strategies and there are efficient algorithms to find them.

\end{itemize}

For more details on how these results follow from the lemmas as well
as formal statements of them, see~\cite{scrip07}.

\section{Technical Appendix}\label{sec:theory}

\lem \label{lem:stationary}
For all states $x$ of the Markov chain whose states are allocations
of money to agents consistent with agent strategies $\vec{k}$ and
whose transition probabilities are determined by $\vec{k}$ and
$(T,\vec{f},n,m)$, let $w_x = \prod_{i} ( \beta_i \chi_i /
\rho_i)^{x_i}$, and let $Z = \sum_{y} w_y$.  Then the limit
distribution of the Markov chain is $\pi_x = w_x / Z$. \elem

\prf
If $S_{xy}$ is the probability of transitioning from state $x$
to state $y$, it is sufficient to show that $\pi_x S_{xy} = \pi_y
S_{yx}$ for all states $x$ and $y$ and that the $\pi_x$ are
non-negative and $\sum_x \pi_x = 1$.  The latter two conditions
follow immediately from the definitions of $w_x$ and $Z$. To check
the first condition, let $x$ and $y$ be adjacent states such that
$y$ is reached from $x$ by $i$ spending a dollar and $j$ earning a
dollar.  This means that for $S_{xy}$ to happen, $i$ must be chosen
to spend a dollar and $j$ must be able to work and chosen to earn
the dollar.  Similarly for $S_{yx}$ to happen, $j$ must be chosen to
spend a dollar and $i$ must be able to work and chosen to earn the
dollar.  All other agents have the same amount of money in each
state, and so will make the same decision in each state.  Thus the
probabilities associated with each transition differ only in the
relative likelihoods of $i$ and $j$ being chosen at each point.  To
be more precise, for some $p$ (which captures the actions of the
other agents), $S_{xy} = \rho_i \beta_j \chi_j p$ and $S_{yx} =
\rho_j \beta_i \chi_i p$.  Also note that since $x$ and $y$ differ
only by $i$ having a dollar in $x$ and $j$ having it is $y$, $\pi_x
= (\beta_i \chi_i \rho_j / \rho_i \beta_j \chi_j) \pi_y$.  Combining
these facts gives:
$$\pi_x S_{xy} = (\frac{\beta_i \chi_i \rho_j}{\rho_i \beta_j \chi_j}
\pi_y)(\frac{\rho_i \beta_j \chi_j}{\rho_j \beta_i \chi_i} S_{yx}) =
\pi_y S_{yx}.$$
\eprf

Note that Lemma 3.1 of \cite{scrip07} which assumes payoff
heterogeneity is now a simple special case of
Lemma~\ref{lem:stationary}.

\cor \label{cor:stationary}
Under the assumption of payoff heterogeneity ($\beta_i = \beta$,
$\chi_i = \chi$, and $\rho_i = \rho$ for all $i$), the limit distribution
is the uniform distribution.
\ecor

\prf
$w_x = (\beta \chi / \rho)^{mn}$ for all $x$, so the limit
distribution is the uniform distribution.
\eprf

To make use of our understanding of the limit distribution from
Lemma~\ref{lem:stationary}, we define a distribution $q$ and a
distribution $M^*_{\vec{f},m}$ that minimizes entropy relative to
it, subject to certain constraints.

\dfn \label{def:q}
Given a system $(T,\vec{f},n,m)$ where agents play strategies $\vec{k}$,
let $\omega_t = \beta_t \chi_t / \rho_t$ for each type $t$.  Let $q$
be the probability distribution on agent types $t$ and amounts of
money $i$
$$q^t_i~=~(\omega_t)^i~/~(\sum_t\sum_{j = 0}^{k_t}~(\omega_t)^j).$$
\edfn

\medskip

\dfn \label{def:minrelent}
Given a system $(T,\vec{f},n,m)$ where agents play strategies
$\vec{k}$,  $M^*$ is the distribution
that minimizes relative entropy $D(M^*||q)$
subject to the constraints
$$\sum_{j=0}^{k_t} (M^*)^t_j = f_t~ \text{ and}$$
$$\sum_{t}\sum_{j=0}^{k_t} j (M^*)^t_j = m.$$
\edfn

\medskip

\lem \label{lem:minrelent}
$$(M^*)^t_i = \frac{f_t \lambda^i q^t_i}
{\sum_{j = 0}^{k_t} \lambda^j q^t_j},$$
where $\lambda$ is chosen so that the second constraint of
Definition~\ref{def:minrelent} is satisfied.
\elem

The proof of Lemma~\ref{lem:minrelent} is omitted because it can be
easily checked using Lagrange multipliers in the manner
of~\cite{jaynes}.

\lem \label{lem:relent}
Given a system $(T,\vec{f},cn,m)$ where agents play strategies
$\vec{k}$, let
$M(\tau)$ be the random variable that gives the
distribution of money in the system at time $\tau$, and let
$M^*$ be defined as in
Definition~\ref{def:minrelent}.  Then for all $\epsilon$, there
exists $n_\epsilon$ such that, if $n > n_\epsilon$, there exists
a time $\tau^*$ such that for all times $\tau > \tau^*$
$$Pr(||M(\tau) - M^*||_2
> \epsilon) < \epsilon.$$
\elem

Note that $|| \cdot ||_2$ is the 2-norm, which
we use for convenience as a measure of distance between distributions.

\prf From Lemma~\ref{lem:stationary}, we know that after a
sufficient amount of time the probability of being in state $x$ will
be close to $\pi_x = w_x / z$.
Therefore, it is sufficient to show that the statement holds in the
limit as $\tau~\rightarrow~\infty$, because if we bound the
probability of distributions being far apart in the limit by
$\epsilon' < \epsilon$, we can take $tau$ large enough that the
additional noise from using only a finite number of steps of the
Markov chain is bounded by $\epsilon - \epsilon'$.
If $M(x)$ is the distribution of money in
state $x$, then we need to show that the set of $x$ such that
$||M(x) - M^*||_2 \geq \epsilon$ has total probability
in the limit distribution of less than $\epsilon$ for sufficiently
large $n$.

Let $D_{\vec{f},m,cn,\vec{k}}$ be the set of distributions achievable
given $\vec{f}$, $m$, $cn$, and $\vec{k}$ and let $d \in
D_{\vec{f},m,cn,\vec{k}}$.
More
precisely $d^t_i$ is the fraction of agents of type $t$ with $i$
dollars.  There may be many states $x$ that have a distribution of
money $d$.  We know from \cite{scrip06} that the number of such
states is proportional to $e^{cnH(d)}$, where $H$ is the standard
entropy function.  Each of these states has the same probability
$w_{d,cn} / z_{cn}$, so the probability of seeing any particular
distribution is proportional to $e^{cnH(d)} w_{d,cn} / z_{cn}$.
First, we show that the distribution that maximizes this
quantity is the distribution $M^*$ from
Definition~\ref{def:minrelent} that minimizes relative entropy.

\begin{eqnarray*}
\argmax_d e^{cnH(d)}w_{d,cn} & = & \argmax_d e^{cnH(d)} \prod_t
\prod_{i = 0}^{k_t} \omega_t^{cn i d^t_i}\\
& = & \argmax_d e^{cnH(d)} \prod_t \prod_{i = 0}^{k_t}
e^{cn i d^t_i \log \omega_t}\\
& = & \argmax_d H(d) + \sum_t \sum_{i = 0}^{k_t}
i d^t_i \log \omega_t\\
& = & \argmax_d \sum_t \sum_{i = 0}^{k_t} [-d^t_i \log d^t_i
+ i d^t_i \log \omega_t]\\
& = & \argmax_d \sum_t \sum_{i = 0}^{k_t}[ -d^t_i \log d^t_i
+ d^t_i \log q^t_i]\\
&= & \argmin_d \sum_t \sum_{i = 0}^{k_t} [d^t_i \log d^t_i
- d^t_i \log q^t_i]\\
&= & \argmin_d \sum_t \sum_{i = 0}^{k_t} d^t_i \log
\frac{d^t_i}{q^t_i}\\
& = & \argmin_d D(d || q).
\end{eqnarray*}

Because all $d \in D_{\vec{f},m,cn,\vec{k}}$ have the property
that $d^t_i cn$ is an integer for all $i$ and $t$, $M^*$ may not be an
element of $ D_{\vec{f},m,cn,\vec{k}}$.  However, for sufficiently
large $cn$ we can find a $d$ arbitrarily close to it.
For a particular $cn$,
take $M^*$ and round each value to the nearest $1 / cn$.
However, the resulting $d$ may have some $t$ for which
$\sum_{i = 0}^{k_t} d^t_i \neq f_t$.
This can be fixed by creating a new $d$ by adding or subtracting $1/cn$
from each $d^t_i$ as appropriate until these constraints are satisfied
(for definiteness, we start with $d^t_0$ for each $t$ and then
proceed in ascending order of $i$ until the constraint for that $t$ is
satisfied).  However, this may still leave the constraint that the
average amount of money is $m$ unsatisfied.  Again, this can be fixed
by slightly adjusting the $d^t_i$, in this case by moving agents from
$d^t_i$ to $d^t_{i-1}$ to subtract money or to $d^t_{i+1}$ to add
money (for definiteness, we start by moving agents downwards from
$d^t_{k_t}$ until it reaches 0 in which case we proceed to $d^t_{k_t -
1}$ if we need to remove money and similarly from $d^t_0$ if we need
to add money).
Our final distance $||d - M^*||_2$ has each term change by
at most $1/cn$ in the first step, $1/cn$ in the second step, and $\sum_t
\sum_{i = 0}^{k_t} 2i / cn$ in the final step (this last is a bound on
how much money could be created or lost by the changes).  This means
each term changes by at most $ (2 |T| (\max_t k_t) + 2) / cn$, which
is $O(1/cn)$.  Thus for a large enough $n_\epsilon$, the resulting $d$
will always be within distance $\epsilon$ of $M^*$.

The remainder of the proof is essentially that of Theorem 3.13 in
\cite{MaxEnt} (though
applied to a different type of constraint).  Let $v_{max}$ be the
maximum value of
$H(d) + \sum_t \sum_{i = 0}^{k_t} i d^t_i \log \omega_t$
at a point in satisfying the
constraints, which we know occurs at $p = M^*$.
We show that there exists
a $v_l$ such that every point not within distance $\epsilon$ of
$p$, the value is at most $v_l$.  Then we
show that there is a point in $D_{\vec{f},m,cn,\vec{k}}$ near $p$ that
has value $v_h > v_l$.

The set of points that for some $n$ are in $D_{\vec{f},m,cn,\vec{k}}$
and are of distance strictly less than $\epsilon$
is an open set.
Therefore its complement is a closed set.
Since this set is closed, there is a point $p_l$ where the value is
maximized in this set.  Take the
value at that point to be $v_l$.
Since entropy is continuous, our function is continuous.
Therefore,
there is some $\epsilon'$ and $v_h$ such that all points within
distance $\epsilon'$ of $p$ have value at least $v_h > v_l$.
The construction of $d$ above allows us find such a point $p_h$.

To complete the proof, we use the fact shown in the proof of Lemma
3.11 of \cite{MaxEnt} that for any $d \in D_{\vec{f},m,cn,\vec{k}}$,
$$\frac{1}{f(cn)} e^{cnH(d)} \leq {cn \choose cnd^{t_1}_0, \ldots,
cnd^{t_1}_{k_{t_1}}, cnd^{t_2}_0, \dots} \leq g(cn) e^{cnH(d)},$$
where $f$ and $g$ are polynomial in $cn$.  We know
that there is some point $p_h$ at which the value is $v_h$.  There
is at least as much total weight near $p$  as there is weight at
$p_h$, so the total weight near $p$ is at least
$$\frac{1}{f(cn) z_{cn}} e^{cn v_h}.$$
The total weight at each point of distance at least $\epsilon$ is at
most
$$\frac{g(cn)}{z_{cn}} e^{cn v_l}.$$
There are at most $(cn+1)^{\sum_t k_t}$ points, so the
total ``bad'' weight is at most
$$\frac{h(cn)}{z_{cn}} e^{cnv_l},$$
where $h(cn) = g(cn) (cn+1)^{\sum_t k_t}$, which is still polynomial in
$n$.  Thus the fraction
of ``bad'' weight is
$$\frac{\frac{h(cn)}{z_{cn}} e^{cn v_l}}{\frac{1}{f(cn) z_{cn}} e^{cn
v_h}} = \frac{f(cn)h(cn)}{e^{cn(v_h - v_l)}}.$$
This is the ratio of a polynomial to an exponential, so the
probability of seeing a distribution of distance greater than
$\epsilon$  from $M^*$ goes to zero as $n$ goes to
infinity.
\eprf

As before, Theorem 3.1 and Corollary 3.1 of~\cite{scrip07}, which
assume payoff
heterogeneity, are a special case of Lemma~\ref{lem:relent}.

\cor \label{cor:relent}
Given a payoff-heterogeneous system with $cn$ agents where a fraction
$\pi_k$ of agents plays strategy $S_k$ and
the average amount of money is $m$,
let $M_{\vec{\pi},n,m}(t)$ be the
the random variable that gives the
distribution of money
in the system at time $t$, and let
$M^*_{\vec{\pi},m}$ be the distribution that maximizes entropy subject
to the constraints imposed by $m$ and $\vec{\pi}$.
Then for all $\epsilon$,
there exists $n_\epsilon$ such that, if $n > n_\epsilon$, there
exists a time $t^*$ such that for all $t > t^*$,
$$\Pr(||M_{\vec{\pi},cn,m}(t) - M^*_{\vec{\pi},m}||_2 > \epsilon) <
\epsilon.$$
Furthermore,
$(M^*_{\vec{\pi},m})^k_i = \pi_k \lambda^i / \sum_{j = 0}^k \lambda^j$
where $\lambda$ is chosen to ensure that the constraint imposed by $m$
is satisfied.
\ecor

\prf
$\omega_t$ has the same value $\omega$ for all types $t$.  Replacing
$\lambda$ by $\omega \lambda$ in the formula for $M^*$ gives a formula
equivalent to Lemma~\ref{lem:minrelent}.  Note that there we were
maximizing entropy and here we are minimizing relative entropy, but
the two are equivalent because when $q$ is the uniform distribution
$D(p||q) = z - H(p)$ where $z$ is a constant~\cite{cover}.
\eprf

\lem \label{lem:monotone}
Let $M_0 = \sum_t (M^*)^t_0$ and let
$M_k = \sum_t (M^*)^t_{k_t}$ (the fraction of
agents with no money and at their threshold respectively).
Suppose each type of agent changes strategy from $k_t$ to $k_t'$
such that $k_t' \geq k_t$ for all $t$.  Then $M_0' \geq M_0$ and
$M_k' \leq M_k$.
\elem

\prf
Since we have a finite set of types, it suffices to consider the case
where a single type $t^*$ increases its strategy by 1.  First, we will
show that
after the change $\lambda$ has decreased.  Suppose that it remained
unchanged.  From the formula in Lemma~\ref{lem:minrelent},
$(M^*)^t_i$ for $t \neq t^*$ would be unchanged.  For
$t^*$, the terms $(M^*)^{t^*}_0$ to
$(M^*)^{t^*}_{k_{t^*}}$ will decrease because there is a
new term in the divisor and the sum of the decreases appears as a new
term $(M^*)^{t^*}_{k_{t^*}}+1$.  This means that the total
amount of money in the system has increased.  Since the total amount
of money is unchanged, $\lambda$ must decrease.

From the formula, it immediately follows that $M_0' \geq M_0$ because
the numerator of each term is effectively unchanged and the
denominator has decreased (the $q^t_i$ have
changed, but the changes cancel out since the
change is to the constant by which each is divided).
For $t \neq t^*$, the derivative of $(M^*)^t_{k_t}$
relative to $\lambda$ is negative, so those terms of $M_k'$ have
decreased.

Finally, we need to show that $(M^*)^t_{k_t} \geq
((M')^*)^t_{k_t + 1}$.  Suppose otherwise.  Then we have

\begin{eqnarray*}
((M')^*)^t_{k_t} & = &\frac{((M')^*)^t_{k_t +
1}}{\lambda' \omega_t}\\
& > &\frac{((M')^*)^t_{k_t + 1}}{\lambda \omega_t}\\
& > &\frac{(M^*)^t_{k_t}}{\lambda \omega_t}\\
& > & (M^*)^t_{k_t - 1}.
\end{eqnarray*}

More generally, this means that $(M^*)^t_i >
((M')^*)^t_{i - 1}$.   But this would mean that $\sum_{i =
0}^{k_t + 1} ((M')^*)^t_{i} > f_t$, which violates a
constraint.
\eprf

\end{document}